\documentclass[aps,prd,nofootinbib,showpacs,superscriptaddress,11pt]{revtex4}
\usepackage{txfonts}
\usepackage{graphicx}
\usepackage{dcolumn}
\usepackage{bm}
\usepackage{amssymb}
\usepackage{latexsym}

\newcommand{\be}{\begin{equation}}
\newcommand{\ee}{\end{equation}}
\newcommand{\bq}{\begin{eqnarray}}
\newcommand{\eq}{\end{eqnarray}}

\begin{document}

\title{Holographic dark energy models: A comparison from the latest observational data}

\author{Miao Li}
\email{mli@itp.ac.cn} \affiliation{Institute of Theoretical Physics,
Chinese Academy of Science, Beijing 100080, China}
\affiliation{Kavli Institute for Theoretical Physics China, Chinese
Academy of Sciences, Beijing 100080, China}
\author{Xiao-Dong Li}
\email{renzhe@mail.ustc.edu.cn} \affiliation{Interdisciplinary
Center for Theoretical Study, University of Science and Technology
of China, Hefei 230026, China} \affiliation{Institute of Theoretical
Physics, Chinese Academy of Science, Beijing 100080, China}
\author{Shuang Wang}
\email{swang@mail.ustc.edu.cn} \affiliation{Department of Modern
Physics, University of Science and Technology of China, Hefei
230026, China} \affiliation{Institute of Theoretical Physics,
Chinese Academy of Science, Beijing 100080, China}
\author{Xin Zhang}
\email{zhangxin@mail.neu.edu.cn} \affiliation{Department of Physics,
College of Sciences, Northeastern University, Shenyang 110004,
China} \affiliation{Kavli Institute for Theoretical Physics China,
Chinese Academy of Sciences, Beijing 100080, China}

\begin{abstract}

The holographic principle of quantum gravity theory has been applied
to the dark energy (DE) problem, and so far three holographic DE
models have been proposed: the original holographic dark energy
(HDE) model, the agegraphic dark energy (ADE) model, and the
holographic Ricci dark energy (RDE) model. In this work, we perform
the best-fit analysis on these three models, by using the latest
observational data including the Union+CFA3 sample of 397 Type Ia
supernovae (SNIa), the shift parameter of the cosmic microwave
background (CMB) given by the five-year Wilkinson Microwave
Anisotropy Probe (WMAP5) observations, and the baryon acoustic
oscillation (BAO) measurement from the Sloan Digital Sky Survey
(SDSS). The analysis shows that for HDE, $\chi_{min}^{2}=465.912$;
for RDE, $\chi_{min}^{2}=483.130$; for ADE,
$\chi_{min}^{2}=481.694$. Among these models, HDE model can give the
smallest $\chi^2_{min}$. Besides, we also use the Bayesian evidence
(BE) as a model selection criterion to make a comparison. It is
found that for HDE, ADE, and RDE, $\Delta \ln \mathrm{BE}= -0.86$,
$-5.17$, and $-8.14$, respectively. So, it seems that the HDE model
is more favored by the observational data.

\end{abstract}


\maketitle

\section{Introduction}\label{sec:intro}

Observations of Type Ia supernovae (SNIa) \cite{Riess}, cosmic
microwave background (CMB) \cite{spergel} and large scale structure
(LSS) \cite{Tegmark} all indicate the existence of mysterious dark
energy (DE) driving the current accelerating expansion of the
universe. The most obvious theoretical candidate of dark energy is
the cosmological constant $\Lambda$, which can fit the observations
well, but is plagued with the fine-tuning problem and the
coincidence problem \cite{Weinberg}. Numerous other dynamical DE
models have also been proposed in the literature, such as
quintessence \cite{quint}, phantom \cite{phantom}, $k$-essence
\cite{k}, tachyon \cite{tachyonic}, hessence \cite{hessence},
Chaplygin gas \cite{Chaplygin}, Yang-Mills condensate \cite{YMC},
ect.

Actually, the DE problem may be in essence an issue of quantum
gravity \cite{Witten}. However, by far, a complete theory of quantum
gravity has not been established, so it seems that we have to
consider the effects of gravity in some effective field theory in
which some fundamental principles of quantum gravity should be taken
into account. It is commonly believed that the holographic principle
\cite{'t Hooft93} is just a fundamental principle of quantum
gravity. Based on the effective quantum field theory, Cohen et al.
\cite{Cohen:1998zx} pointed out that the quantum zero-point energy
of a system with size $L$ should not exceed the mass of a black hole
with the same size, i.e. $L^3\rho_{vac}\leq LM_{Pl}^2$, where
$\rho_{vac}$ is the quantum zero-point energy density, and
$M_{Pl}\equiv 1/\sqrt{8\pi G}$ is the reduced Planck mass. This
observation relates the ultraviolet (UV) cutoff of a system to its
infrared (IR) cutoff. When we take the whole universe into account,
the vacuum energy related to this holographic principle can be
viewed as dark energy (its energy density is denoted as $\rho_{de}$
hereafter). The largest IR cutoff $L$ is chosen by saturating the
inequality, so that we get the holographic dark energy density
\begin{equation}
\rho_{de}=3c^2 M_{Pl}^2L^{-2}~,\label{de}
\end{equation}
where $c$ is a numerical constant characterizing all of the
uncertainties of the theory, and its value can only be determined by
observations. If we take $L$ as the size of the current universe,
for instance the Hubble radius $H^{-1}$, then the dark energy
density will be close to the observational result. However, Hsu
\cite{Hsu} pointed out this yields a wrong equation of state (EOS) for DE.

Li \cite{Li} suggested to choose the future event horizon of the
universe as the IR cutoff. This is the original holographic dark
energy model. Subsequently, Cai proposed \cite{Cai} that the age of
the universe can be chosen as the IR cutoff, and this model is
called agegraphic dark energy model. A new version of this model
replacing the age of the universe by the conformal age of the
universe was also discussed \cite{Wei}, in order to avoid some
internal inconsistencies in the original model. Furthermore, Gao et
al. \cite{Gao} proposed to consider the average radius of the Ricci
scalar curvature as the IR cutoff, and this model is called
holographic Ricci dark energy model. (For convenience, hereafter we
will call them HDE, ADE, and RDE, respectively.) Although these
three models have been studied widely  \cite{refHDE,refADE,refRDE},
so far no comparison of them has been made. It would be very
interesting to constrain the holographic dark energy models
by using the latest observational data, and then make a comparison
for them by using the proper model selection criterion. This will be
done in this work.

This paper is organized as follows: In Section 2, we briefly review
the holographic DE models. In Section 3, we present the method of
data analysis, as well as the model selection criterion. In Section
4, we show the results of the cosmological constraints, and we also
use the Bayesian evidence (BE) to make a comparison. Section 5 is a
short summary. In this work, we assume today's scale factor
$a_{0}=1$, so the redshift $z$ satisfies $z=a^{-1}-1$; the subscript
``0'' always indicates the present value of the corresponding
quantity, and the unit with $c=\hbar=1$ is used.

\section{Models}

In this section, we shall briefly review the holographic
DE models. For a spatially flat (the assumption of flatness
is motivated by the inflation scenario) Friedmann-Robertson-Walker
(FRW) universe with matter component $\rho_{m}$ and dark energy
component $\rho_{de}$, the Friedmann equation reads
\begin{equation}\label{Fri}
3M_{Pl}^2H^2=\rho_{m}+\rho_{de}~,
\end{equation}
or equivalently,
\begin{equation}
E(z)\equiv {H(z)\over H_0}=\left(\Omega_{m0}(1+z)^3\over
1-\Omega_{de}\right)^{1/2},\label{Ez}
\end{equation}
where $H\equiv \dot{a}/a$ is the Hubble parameter, $\Omega_{m0}$ is
the present fractional matter density, and $\Omega_{de}\equiv
\frac{\rho_{de}}{\rho_{c}} = \frac{\rho_{de}}{3M_{Pl}^2H^2}$ is the
fractional dark energy density. In this work, for simplicity, we
shall not discuss the issue of interaction between dark matter and
dark energy, so we have
\begin{equation}\label{m}
\dot{\rho}_{m}+3H\rho_{m}=0~,
\end{equation}
\begin{equation}\label{e}
\dot{\rho}_{de}+3H(1+w_{de})\rho_{de}=0~,
\end{equation}
where the over dot denotes the derivative with respect to the cosmic time $t$,
and $w_{de}$ is the EOS of DE.

\subsection{The HDE model}

For this model, the IR cutoff is chosen as the future event horizon
of the universe,
\begin{equation}
L = a\int_{t}^{\infty}\frac{dt^{\prime}}{a}
=a\int_{a}^{\infty}\frac{da^{\prime}}{Ha^{\prime2}}~.\label{rh}
\end{equation}
Taking derivative for Eq. (\ref{de}) with respect to $x=\ln a$ and
making use of Eq. (\ref{rh}), we get
\begin{equation}
\rho _{de}^{\prime}\equiv \frac{d\rho _{de}}{dx} = 2 \rho
_{de}(\frac{\sqrt{\Omega _{de}}}{c}-1).  \label{Derivative}
\end{equation}
Combining Eqs.~(\ref{e}) and (\ref{Derivative}), we obtain the EOS
for HDE,
\begin{equation}
w_{de}=-{1\over 3}-{2\over 3c}\sqrt{\Omega_{de}}.\label{whde}
\end{equation}
Directly taking derivative for $\Omega_{de}={c^2/ (H^2L^2)}$, and
using Eq. (\ref{rh}), we get
\begin{equation}
\Omega_{de}'=2\Omega_{de}\left(\epsilon-1+{\sqrt{\Omega_{de}}\over
c}\right),\label{deeps}
\end{equation}
where $\epsilon\equiv -{\dot{H}/ H^2}=-{H'/ H}$. From
Eqs.~(\ref{Fri}), (\ref{m}), (\ref{e}), and (\ref{whde}), we have
\begin{equation}
\epsilon={3\over 2}(1+w_{de}\Omega_{de})={3\over
2}-{\Omega_{de}\over 2}-{\Omega_{de}^{3/2}\over c},
\end{equation}
for this case. So, we have the equation of motion, a differential
equation, for $\Omega_{de}$,
\begin{equation}
\Omega _{de}^{\prime}=\Omega _{de}(1-\Omega
_{de})\left(1+\frac{2}{c}\sqrt{\Omega _{de}}\right).  \label{Density
1}
\end{equation}
Since $\frac{d}{dx}=-(1+z)\frac{d}{dz}$, we get
\begin{equation}
\frac{d
\Omega_{de}}{dz}=-(1+z)^{-1}\Omega_{de}(1-\Omega_{de})\left(1+{2\over
c}\sqrt{\Omega_{de}}\right).  \label{Density 2}
\end{equation}
Solving numerically Eq .~(\ref{Density 2}) and substituting the
corresponding results into  Eq .~(\ref{Ez}), the key function $E(z)$
can be obtained. It should be mentioned that there are two model
parameters, $\Omega_{m0}$ and $c$, in the HDE model.

\subsection{The ADE model}

Since there are some internal inconsistencies in the original ADE
model, we will discuss the new version of ADE model which suggests
to choose the conformal age of the universe
\begin{equation}
\eta \equiv \int \frac{dt}{a}=\int \frac{da}{a^{2}H}~,\label{etaq}
\end{equation}
as the IR cutoff, so the density of ADE is
\begin{equation}
\rho_{de}=3n^{2}M_{Pl}^{2}\eta^{-2}.\label{rhoq}
\end{equation}
To distinguish from the HDE model, a new constant parameter $n$ is
used to replace the old parameter $c$. Taking derivative for Eq.
(\ref{rhoq}) with respect to $x$ and making use of Eq. (\ref{etaq}),
we get
\begin{equation}
\rho _{de}^{\prime}= -2 \rho_{de}\frac{\sqrt{\Omega _{de}}}{n a}.  \label{Derivative2}
\end{equation}
This means that the EOS of ADE is
\begin{equation}
w_{de}=-1+{2\over 3n}{\sqrt{\Omega_{de}}\over a}.\label{wade}
\end{equation}
Taking derivative for $\Omega_{de}=n^2/(H^2\eta^2)$, and considering
Eq.~(\ref{etaq}), we obtain
\begin{equation}
\Omega'_{de}=2\Omega_{de}\left(\epsilon-{\sqrt{\Omega_{de}}\over
na}\right).
\end{equation}
In this case, we have
\begin{equation}
\epsilon={3\over 2}(1+w_{de}\Omega_{de})={3\over 2}-{3\over
2}\Omega_{de}+{\Omega_{de}^{3/2}\over na}.
\end{equation}
Hence, we get the equation of motion for $\Omega _{de}$,
\begin{equation}
\Omega _{de}^{\prime}=\Omega_{de}(1-\Omega_{de})\left(3-\frac{2}{n}\frac{\sqrt{\Omega_{de}}}{a}\right)~,
\end{equation}
and this equation can be rewritten as
\begin{equation}
\frac{d \Omega_{de}}{dz}=-\Omega_{de}(1-\Omega_{de})\left(3(1+z)^{-1}-\frac{2}{n}\sqrt{\Omega_{de}}\right)~\label{keyq}.
\end{equation}
As in \cite{Wei}, we choose the initial condition,
$\Omega_{de}(z_{ini})=n^2(1+z_{ini})^{-2}/4$, at $z_{ini}=2000$,
then Eq. (\ref{keyq}) can be numerically solved. Substituting the
results of Eq. (\ref{keyq}) into  Eq .~(\ref{Ez}), the key function
$E(z)$ can be obtained. Notice that once $n$ is given,
$\Omega_{m0}=1-\Omega_{de}(z=0)$ can be natural obtained by solving
Eq.(\ref{keyq}), so the ADE model is a single-parameter model.

\subsection{The RDE model}
For a spatially flat FRW universe, the Ricci scalar is
\begin{equation}
{\cal R}=-6\left(\dot{H}+2H^2\right).
\end{equation}
As suggested by Gao et al. \cite{Gao}, the energy density of DE is
\begin{equation}
\rho_{de}={3\alpha\over 8\pi
G}\left(\dot{H}+2H^2\right)=-{\alpha\over 16\pi G}{\cal R},
\end{equation}
where $\alpha$ is a positive numerical constant to be determined by
observations. Comparing to Eq. (\ref{de}), it is seen that if we
identify the IR cutoff $L$ with $-{\cal R}/6$, we have $\alpha=c^2$.
As pointed out by Cai et al. \cite{Cai2}, the RDE can be viewed as
originated from taking the causal connection scale as the IR cutoff
in the holographic setting. Now the Friedmann equation can be
written as
\begin{equation}
H^2={8\pi G\over 3}\rho_{m0}e^{-3x}+\alpha\left({1\over 2}{dH^2\over dx}+2H^2\right),
\end{equation}
and this equation can be further rewritten as
\begin{equation}
E^2=\Omega_{m0}e^{-3x} +\alpha\left({1\over 2}{dE^2\over dx}+2E^2\right),
\end{equation}
where $E\equiv H/H_{0}$. Solving this equation, and using the
initial condition $E_0=E(t_0)=1$, we have
\begin{equation}
E(z)=\left(\frac{2
\Omega_{m0}}{2-\alpha}(1+z)^{3}+(1-{2\Omega_{m0}\over 2 -\alpha})
(1+z)^{(4-{2\over\alpha})}\right)^{1/2}.\label{Ea}
\end{equation}
There are also two model parameters, $\Omega_{m0}$ and $\alpha$, in
RDE model.

\section{Methodology}

\subsection{Data analysis}
In the following, we constrain the model parameters of these three
DE models by using the latest observational data including the
Union+CFA3 sample of 397 SNIa, the shift parameter of the CMB given
by the five-year Wilkinson Microwave Anisotropy Probe (WMAP5)
observations, and the baryon acoustic oscillation (BAO) measurement
from the Sloan Digital Sky Survey (SDSS).

First, we consider the latest 397 Union+CFA3 SNIa data, the distance
modulus $\mu_{ obs}(z_i)$, compiled in Table 1 of \cite{Hicken}.
This dataset add the CFA3 sample to the 307 Union sample
\cite{Kowalski}. The theoretical distance modulus is defined as
\begin{equation}
\mu_{th}(z_i)\equiv 5 \log_{10} {D_L(z_i)} +\mu_0,
\end{equation}
where $\mu_0\equiv 42.38-5\log_{10}h$ with $h$ the Hubble constant
$H_0$ in units of 100 km/s/Mpc, and
\begin{equation}
D_L(z)=(1+z)\int_0^z {dz'\over E(z';{\bf \theta})}
\end{equation}
is the Hubble-free luminosity distance $H_0d_L$ (here $d_L$ is the
physical luminosity distance) in a spatially flat FRW universe, and
here ${\bf \theta}$ denotes the model parameters. The $\chi^2$ for
the SNIa data is
\begin{equation}
\chi^2_{SN}({\bf\theta})=\sum\limits_{i=1}^{397}{[\mu_{obs}(z_i)-\mu_{th}(z_i)]^2\over \sigma_i^2},\label{ochisn}
\end{equation}
where $\mu_{obs}(z_i)$ and $\sigma_i$ are the observed value and the
corresponding 1$\sigma$ error of distance modulus for each
supernova, respectively. The parameter $\mu_0$ is a nuisance
parameter but it is independent of the data and the dataset.
Following \cite{Nesseris}, the minimization with respect to $\mu_0$
can be made trivially by expanding the $\chi^2$ of Eq.
(\ref{ochisn}) with respect to $\mu_0$ as
\begin{equation}
\chi^2_{SN}({\bf\theta})=A({\bf\theta})-2\mu_0
B({\bf\theta})+\mu_0^2 C,
\end{equation}
where
\begin{equation}
A({\bf\theta})=\sum\limits_{i=1}^{307}{[\mu_{obs}(z_i)-\mu_{th}(z_i;\mu_0=0,{\bf\theta})]^2\over \sigma_i^2},
\end{equation}
\begin{equation}
B({\bf\theta})=\sum\limits_{i=1}^{307}{\mu_{obs}(z_i)-\mu_{th}(z_i;\mu_0=0,{\bf\theta})\over \sigma_i^2},
\end{equation}
\begin{equation}
C=\sum\limits_{i=1}^{307}{1\over \sigma_i^2}.
\end{equation}
Evidently, Eq. (\ref{ochisn}) has a minimum for $\mu_0=B/C$ at
\begin{equation}
\tilde{\chi}^2_{
SN}({\bf\theta})=A({\bf\theta})-{B({\bf\theta})^2\over C}.\label{tchi2sn}
\end{equation}
Since $\chi^2_{SN, min}=\tilde{\chi}^2_{SN, min}$, instead
minimizing $\chi^2_{SN}$ one can minimize $\tilde{\chi}^2_{SN}$ which is independent of the nuisance parameter $\mu_0$.

Next, we consider the constraints from CMB and LSS observations. For
the CMB data, we use the CMB shift parameter $R$; for the LSS data,
we use the parameter $A$ of the BAO measurement. It is widely
believed that both $R$ and $A$ are nearly model-independent and
contain essential information of the full CMB and LSS BAO data. The
shift parameter $R$ is given by \cite{Bond,Wangyun}
\begin{equation}
R\equiv \Omega_{m0}^{1/2}\int_0^{z_{rec}}{dz'\over E(z')},
\end{equation}
where the redshift of recombination $z_{rec}=1090$ \cite{Komatsu}.
The shift parameter $R$ relates the angular diameter distance to the
last scattering surface, the comoving size of the sound horizon at
$z_{rec}$ and the angular scale of the first acoustic peak in CMB
power spectrum of temperature fluctuations \cite{Bond,Wangyun}. The
measured value of $R$ has been updated to be $R_{obs}=1.710\pm
0.019$ from WMAP5 \cite{Komatsu}. The parameter $A$ from the
measurement of the BAO peak in the distribution of SDSS luminous red
galaxies is defined as \cite{Eisenstein}
\begin{equation}
A\equiv \Omega_{m0}^{1/2} E(z_{b})^{-1/3}\left[{1\over z_{b}}\int_0^{z_{b}}{dz'\over E(z')}\right]^{2/3},
\end{equation}
where $z_{b}=0.35$. The SDSS BAO measurement \cite{Eisenstein} gives
$A_{obs}=0.469\,(n_s/0.98)^{-0.35}\pm 0.017$, where the scalar
spectral index is taken to be $n_s=0.960$ as measured by WMAP5
\cite{Komatsu}. Notice that both $R$ and $A$ are independent of
$H_0$, thus these quantities can provide robust constraint as
complement to SNIa data on the holographic dark energy models. The
total $\chi^2$ is given by
\begin{equation}
\chi^2=\tilde{\chi}_{SN}^2+\chi_{CMB}^2+\chi_{BAO}^2~,
\end{equation}
where $\tilde{\chi}_{ SN}^2$ is given by Eq. (\ref{tchi2sn}), and
the latter two terms are defined as
\begin{equation}\label{chiCMB}
\chi^2_{CMB}=\frac{(R-R_{obs})^2}{\sigma_R^2},
\end{equation}
and
\begin{equation}\label{chiLSS}
\chi^2_{BAO}=\frac{(A-A_{obs})^2}{\sigma_A^2},
\end{equation}
where the corresponding $1\sigma$ errors are $\sigma_R=0.019$ and
$\sigma_A=0.017$, respectively. As usual, assuming the measurement
errors are Gaussian, the likelihood function
\begin{equation}\label{likelihood}
{\cal{L } } \propto e^{-\chi^2/2}.
\end{equation}
The model parameters yielding a minimal $\chi^{2}$ and a maximal
${\cal{L } }$ will be favored by the observations.

\subsection{Model Comparison}

For comparing different models, a statistical variable must be
chosen. The $\chi _{min}^{2}$ is the simplest one which is widely
used. However, for models with different number of parameters, the
comparison using $\chi^2$ may not be fair, as one would expect that
a model with more parameters tends to have a lower $\chi^2$. In this
work, we will use the BE as a model selection criterion. The BE of a
model $M$ takes the form
\begin{equation}
\mathrm{BE}=\int {\mathcal{L}(\mathbf{d|\theta
},M)\mathbf{p}(\mathbf{\theta }|M)d\mathbf{\theta }},  \label{BE}
\end{equation}
where $\mathcal{L}(\mathbf{d|\theta },M)$ is the likelihood function
given by the model $M$ and parameters $\mathbf{\theta }$
($\textbf{d}$ denotes the data), and $\mathbf{p}(\mathbf{\theta
}|M)$ is the priors of parameters. It is clear that a model favored
by the observations should give a large BE. Since BE is the average
of the likelihood of a model over its prior of the parameter space
and automatically includes the penalties of the number of parameters
and data, it is more reasonable and unambiguous than the $\chi
_{min}^{2}$ in model selection \cite {Liddle}. The logarithm of BE
can be used as a guide for model comparison \cite {Liddle}, and we
choose the $\Lambda $CDM as the reference model: $\Delta \ln
\mathrm{BE}=\ln \mathrm{BE}_{model}-\ln \mathrm{BE}_{\Lambda CDM}$.

\section{Results}

\begin{figure}
\centerline{\includegraphics[width=8cm]{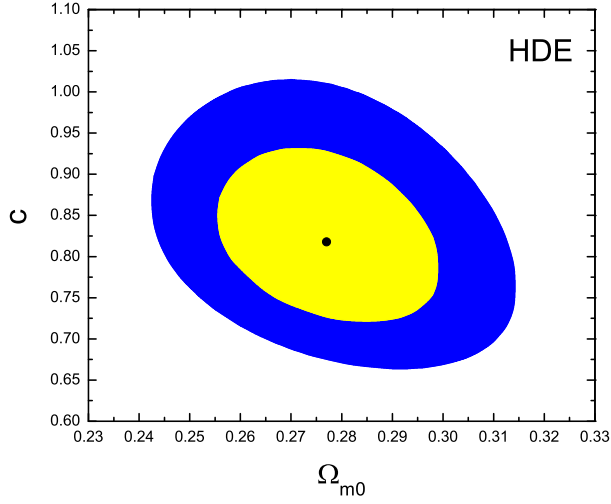}}
\caption{\label{fig1} Probability contours at $68.3\%$ and $95.4\%$
confidence levels in $\Omega_{m0}-c$ plane, for the HDE model. The
round point denotes the best-fit values, $\Omega_{m0}=0.277$ and
$c=0.818$, corresponding to $\chi^2_{min}=465.912$.}
\end{figure}

We will show the results in this section. In Fig.\ref{fig1}, we plot
the contours of $68.3\%$ and $95.4\%$ confidence levels in the
$\Omega_{m0}-c$ plane, for the HDE model. The best-fit model
parameters are $\Omega_{m0}=0.277$ and $c=0.818$, corresponding to
$\chi^2_{min}=465.912$. For $68.3\%$ confidence level,
$\Omega_{m0}=0.277^{+0.022}_{-0.021}$, and
$c=0.818^{+0.113}_{-0.097}$; for $95.4\%$ confidence level,
$\Omega_{m0}=0.277^{+0.037}_{-0.034}$, and
$c=0.818^{+0.196}_{-0.154}$.

\begin{figure}
\centerline{\includegraphics[width=8cm]{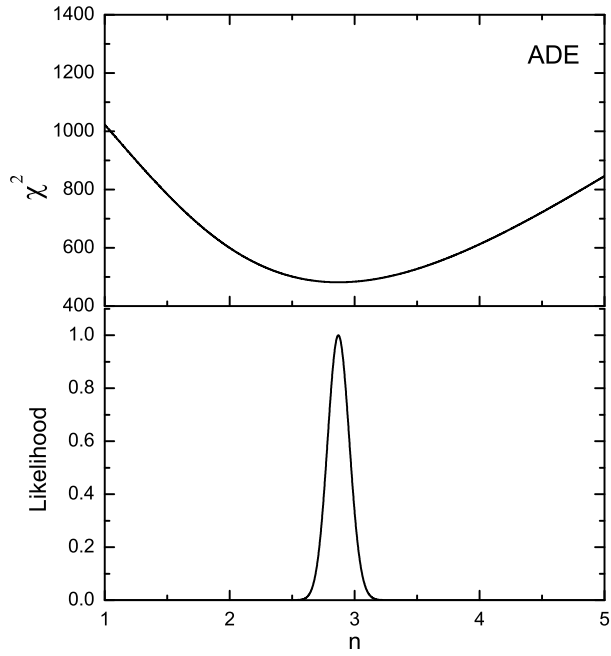}}
\caption{\label{fig2} The $\chi^{2}$ and the corresponding
likelihood ${\cal{L } }$ of the ADE model.}
\end{figure}

\begin{figure}
\centerline{\includegraphics[width=8cm]{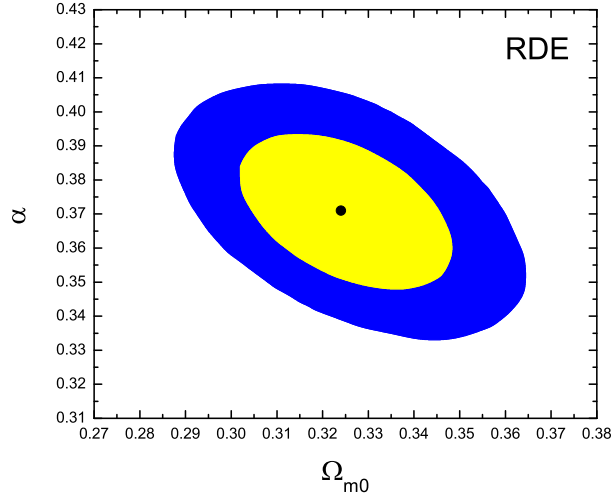}}
\caption{\label{fig3} Probability contours at $68.3\%$ and $95.4\%$
confidence levels in $\Omega_{m0}-\alpha$ plane, for the RDE model.
The round point denotes the best-fit values, $\Omega_{m0}=0.324$ and
$\alpha=0.371$, corresponding to $\chi^2_{min}=483.130$.}
\end{figure}

\begin{figure}
\centerline{\includegraphics[width=8cm]{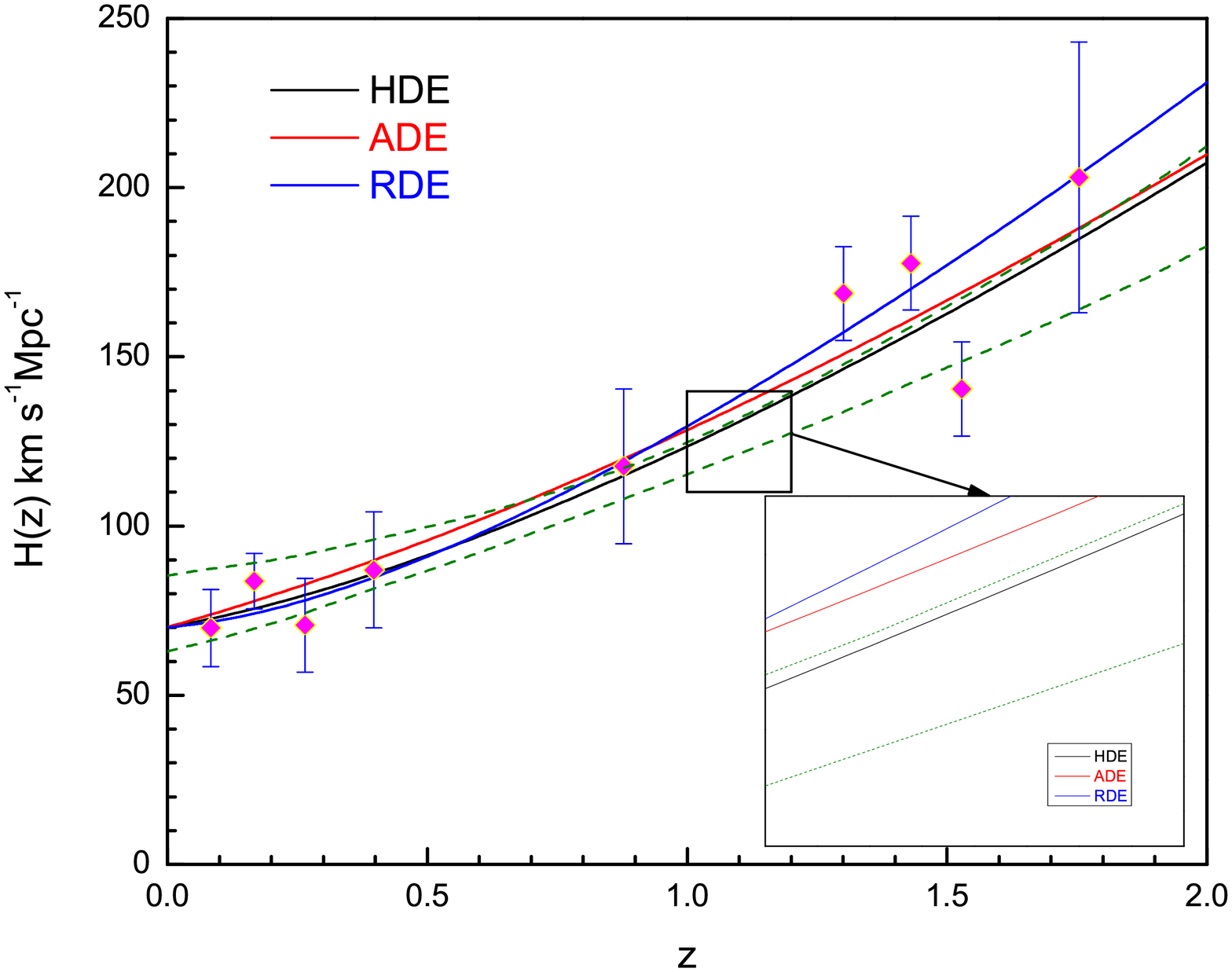}}
\caption{\label{fig4} Comparison of the observed $H(z)$, as square
dots, with the predictions from the holographic DE models. }
\end{figure}

In Fig.\ref{fig2}, we plot the $\chi^{2}$ and the corresponding
likelihood ${\cal{L } }$ of the ADE model. The best-fit model
parameter is $n=2.807$, giving $\chi^2_{min}=481.694$. For $68.3\%$
confidence level, $n=2.807^{+0.087}_{-0.086}$; for $95.4\%$
confidence level, $n=2.807^{+0.176}_{-0.170}$.

In Fig.\ref{fig3}, we plot the contours of $68.3\%$ and $95.4\%$
confidence levels in the $\Omega_{ m0}-\alpha$ plane, for the RDE
model. The best-fit model parameters are $\Omega_{m0}=0.324$ and
$\alpha=0.371$, corresponding to $\chi^2_{min}=483.130$. For
$68.3\%$ confidence level, $\Omega_{m0}=0.324^{+0.024}_{-0.022}$,
and $\alpha=0.371^{+0.023}_{-0.023}$; for $95.4\%$ confidence level,
$\Omega_{m0}=0.324^{+0.040}_{-0.036}$, and
$\alpha=0.371^{+0.037}_{-0.038}$.

Therefore, among these three holographic DE models, the HDE model
can give the smallest $\chi^2_{min}$. As a comparison, we also fit
the $\Lambda$CDM model to the same observational data, and find that
the minimal $\chi^2_{min}=467.775$ for the best-fit parameter
$\Omega_{m0}=0.274$. Note that both HDE model and RDE model are
two-parameter model, while ADE model is a single-parameter model.
So, it is not suitable to choose the $\chi _{min}^{2}$ as a model
selection criterion. In Table \ref{BE}, we list the results of
$\Delta \ln \mathrm{BE}$ for three holographic DE models. It is seen
that although HDE model performs a little poorer than $\Lambda$CDM
model, it performs better than ADE model and RDE model. Therefore,
among these three holographic DE models, HDE is more favored by the
observational data.

\begin{table}
\caption{The results of $\Delta \ln \mathrm{BE}$ for three
holographic DE models.}
\begin{center}
\label{BE}
\begin{tabular}{cccc}
  \hline\hline
  ~~~Model~~~ & ~~~HDE~~~ & ~~~ADE~~~ & ~~~RDE~~~ \\
  \hline
  ~~~$\Delta \ln \mathrm{BE}$~~~ & ~~~$-0.86$~~~ & ~~~$-5.17$~~~ & ~~~$-8.14$~~~ \\
  \hline\hline
\end{tabular}
\end{center}
\end{table}

In Fig.\ref{fig4}, we further compare the observed expansion rate
$H(z)$ \cite{Simon} with that predicted by these three holographic
DE models (for each model, the best-fit values determined by
SNIa+CMB+BAO analysis are taken). Notice that the area surrounded by
two dashed lines shows the $68\%$ confidence interval
\cite{spergel}, and a DE model would be disfavored by the
observation if it gives a curve of $H(z)$ that falls outside this
area. It is seen that among these three holographic DE models, only
the curve of $H(z)$ predicted by the HDE model falls inside this
confidence interval. This result verifies our conclusion, from
another perspective.

\section{Summary}

In this work, we perform the best-fit analysis on three holographic
DE models, by using the latest observational data including the
Union+CFA3 sample of 397 SNIa, the shift parameter of the CMB given
by the WMAP5 observations, and the BAO measurement from the SDSS.
The analysis shows that for HDE, $\chi_{min}^{2}=465.912$; for RDE,
$\chi_{min}^{2}=483.130$; for ADE, $\chi_{min}^{2}=481.694$. Among
these three DE models, the HDE model can give the smallest
$\chi^2_{min}$, which is even smaller than that given by the
$\Lambda$CDM model. Moreover, we use the BE as a model selection
criterion to make a comparison. Although HDE model performs a little
poorer than $\Lambda$CDM model, it performs better than ADE model
and RDE model. Therefore, among these three DE models, HDE is more
favored by the observational data. Finally, adopting the best-fit
values determined by SNIa+CMB+BAO analysis, we compare the observed
expansion rate $H(z)$ with that predicted by these three holographic
DE models. It is found that only the curve of $H(z)$ predicted by
HDE model can fall inside the $68\%$ confidence interval. This
verifies our conclusion from another perspective.

It would be interesting to give tighter constrains on DE models by
adding other cosmological observations, such as gamma-ray bursts
\cite{GRB}, Chandra x-ray observation \cite{X-ray}, and some old
high-redshift objects \cite{Age}. This will be studied in a future
work.

\section*{Acknowledgements}
We would like to thank Hao Wei, Yan Gong, Chang-Jun Gao, and
Feng-Quan Wu, for helpful discussions. This work is supported by the
Natural Science Foundation of China.

%%%%%%%%%%%%%%%%%%%%%%%%%%%%%%%%%%%%%%%

\end{document}